\documentclass{ws-procs975x65}

\begin{document}

\title{ASYMPTOTICALLY ANTI DE SITTER SPACETIMES IN THREE DIMENSIONS}

\author{Luca Bombelli and Arif Mohd$^*$ }

\address{Department of Physics And Astronomy, University of Mississippi,\\
University, MS 38677, U.S.A.\\
$^*$E-mail: arif@phy.olemiss.edu}

\begin{abstract}
We revisit the asymptotically Anti de Sitter spacetimes in three dimensions. Using the conformal-completion technique, we formulate the boundary conditions in a covariant fashion and construct the global charges associated with the asymptotic symmetries. The charges so constructed are conserved for the asymptotic Killing vectors fields but are not conserved for the asymptotic conformal Killing vector fields. The quantity integrated to obtain the global charge is interpreted as the Brown-York boundary stress-energy tensor and it is found not to be traceless. The trace is interpreted as the Trace Anomaly and it turns out to be the same as the Brown-Henneaux central charge. 
\end{abstract}


\bodymatter

\section{Introduction}\label{aba:sec1}
The spacetimes which asymptote to Anti de Sitter  have attracted much attention in the context of the conjectured AdS/CFT correspondence. One (of the many) non-trivial test(s) of the correspondence is provided by an exact matching of the anomalies calculated on the gravity side on one hand and the gauge theory side on the other. Traditionally, the anomalies in the gravity side are calculated by regularizing the gravitational action by subtracting some counterterms constructed out of the boundary-covariant quantities \cite{Balasubramanian:1999re}. This regularization depends upon the representative of the conformal class of the metric on the boundary and hence breaks conformal invariance thus leading to the conformal anomaly in the odd dimensional spacetimes \cite{Henningson:1998gx}. However,  if one constructs the conserved quantities using only the conformal structure of the boundary, working directly with the equations of motion instead of the action, one doesn't find any anomalies in any dimensions greater than three \cite{Ashtekar:1999jx}.  Now, we know from the work of  Brown and Henneaux that in three dimensions the poisson bracket algebra of the canonical generators of the asymptotic symmetries is the central extension of the lie algebra of the asymptotic symmetries \cite{Brown:1986nw}. The anomaly shows up as the central charge. The purpose of this article is to investigate if we can reproduce this anomaly using the conformal completion techniques of Ref. ~\refcite{Ashtekar:1999jx} in three dimensions.

\section{Boundary Conditions, Global Charges and Anomaly}

A 2+1 dimensional spacetime $(\widetilde{M},\widetilde{g}_{ab})$ will be called Asymptotically AdS if there exists a manifold $(M,g_{ab})$ with boundary $\mathcal{I}$ and a diffeomorphism from $\widetilde{M}$ onto $M-\mathcal{I}$ such that the following conditions hold :
\begin{romanlist}
\item $\exists$ a function $\Omega$ on $M$ such that on $\widetilde{M}$ 
\begin{align*}
g_{ab}= \Omega^{2} \widetilde{g}_{ab}.
\end{align*}
\item $\mathcal{I}$ is topologically $\mathbb{S}^1\times\mathbb{R}$,	$\Omega$ vanishes on $\mathcal{I}$ while $n_a :=\nabla_a \Omega$ is nowhere vanishing on $\mathcal{I}$.
\item on $\widetilde{M}$, $\widetilde{g}_{ab}$ satisfies
\begin{align*}
\widetilde{R}_{ab} - \frac{1}{2} R \widetilde{g}_{ab} + \Lambda \widetilde{g}_{ab}= 8\pi G T_{ab}.
\end{align*}
\item $\Omega^{-1}T_{ab}$ admits a smooth limit on $\mathcal{I}$.
\end{romanlist}	

By choosing a different conformal factor $\Omega^\prime = \omega \Omega $, we can show that 
\begin{align*}
\nabla^\prime_a n^\prime_b = \omega \nabla_a n_b + g_{ab} n^c \nabla_c \omega,  
\end{align*}
where, $n^\prime_a = \nabla_a (\omega \Omega)$, $\nabla^\prime$ is the covariant derivative with respect to the metric rescaled by $(\omega \Omega)^2$ and the equality holds on $\mathcal{I}$.
Therefore, we can choose the conformal factor such that $\nabla_a n_b$ vanishes on $\mathcal{I}$. We suppose that we have made this choice. In particular, the extrinsic curvature of $\mathcal{I}$ vanishes. Notice that $ \omega $ itself is unrestricted on $\mathcal{I}$. \par
Writing the Einstein equation in terms of the unphysical geometrical quantities but the physical matter we get,
\begin{align}
S_{ab}+\frac{1}{\Omega}\nabla_a n_b - \frac{1}{\Omega^2}g_{ab}(n.n)-\frac{1}{2\Omega^2}g_{ab}=L_{ab}
\label{EinsteinEqn}
\end{align} 
where,
\begin{align}
S_{ab}&= R_{ab}- \frac{1}{4}R g_{ab} ,\\
L_{ab}&=8 \pi G (T_{ab}-\frac{1}{2}T g_{ab}).
\end{align}
Mutliplying Eq. (\ref{EinsteinEqn})  by $\Omega^2$ and taking the limit $\Omega \rightarrow 0 $ we get,
 \begin{align}
 n.n = - \Lambda \equiv \frac{1}{\ell^2} \text{\hspace{2mm}on\hspace{1mm}} \mathcal{I}.
 \end{align}
 Hence, the boundary $\mathcal{I}$ is time-like. 
Taking the antisymmetrized derivative of Eq. (\ref{EinsteinEqn})  we get,
\begin{align}
\nabla_{[c}S_{a]b} = \Omega^{-1} \left ( \nabla_{[c}(\Omega L_{a]b})-g_{b[c}L_{a]d}n^d \right ).
\end{align}

Finally, taking the limit $\Omega \rightarrow 0 $ and the appropriate projections on $\mathcal{I}$ we get a Differential \textquotedblleft Conservation\textquotedblright Law,
\begin{align}
D_{\alpha}\tau^{\alpha}_{\beta} = 8 \pi G \hat{T}_{\alpha \gamma} n^{\alpha}h^{\gamma}_{\beta}
\label{DifferentialConservationLaw}
\end{align}
where, $\tau_{\alpha \beta}$ is $S_{ab}$ with the indices projected on the boundary, $\mathcal{I}$. We interpret $\dfrac{\ell}{8\pi G} \tau_{\alpha \beta}$ as the Brown-York boundary stress-energy tensor \cite{Brown:1992br}. Given a vector field $\xi^\beta$ on the boundary, we integrate $D_{\alpha}(\tau^{\alpha}_{\beta} \xi^\beta)$ over a portion of the boundary $\Delta \mathcal{I}$  enclosed between the cross-sections $C_1$ and $C_2$ to get the integrated version of the \textquotedblleft Conservation\textquotedblright Law:
\begin{align}
\frac{\ell}{8 \pi G} \oint_{C_1-C_2}\tau_{\alpha  \beta} \xi ^\alpha dS^\beta = \int_{\Delta \mathcal{I}} \hat{T}_{\alpha \gamma} \xi^\alpha ds^\gamma + \frac{1}{2}\int_{\Delta \mathcal{I}} \mathcal{A} D_\lambda \xi^\lambda 
\label{IntegralConservationLaw}
\end{align} 
where,   $\mathcal{A}$ is the Trace Anomaly, 
\begin{align}
\mathcal{A}= - \frac{\ell}{8 \pi G} \frac{^{(2)}R}{2}.
\end{align}
Comparing $\mathcal{A}$ with the trace anomaly in the two dimensional conformal field theory, $T^\mu_\mu = -\dfrac{cR}{24 \pi}$, we get the Brown-Henneaux central charge  \cite{Brown:1986nw} ,
\begin{align}
c = \frac{3\ell}{2G}.
\end{align}
The global charge corresponding to $\xi^a$ is defined as
\begin{align}
 \mathcal{Q}_\xi = \dfrac{\ell}{8 \pi G} \oint_{C}\tau_{\alpha  \beta} \xi ^\alpha ds^\beta
\end{align} 
 From Eq. (\ref{IntegralConservationLaw}) we see that $\mathcal{Q}_\xi$ is conserved for the killing vector(KV) field of $\mathcal{I}$ but is not conserved for the conformal killing vector (CKV) field of $\mathcal{I}$. Although, CKV fields of $\mathcal{I}$ are the generators of the asymptotic symmetry group because they keep the boundary conditions invariant, we see that the conserved quantities correspond only to the KV fields of $\mathcal{I}$. Even in the absence of the matter, the non-conservation of global charges corresponding to the CKV fields is  attributed solely to the anomaly.
\begin{align}
\mathcal{Q}_\xi |_{C_1} - \mathcal{Q}_\xi |_{C_2} = \text{Matter Flux + Anomaly Flux}.
\end{align}
 We have therefore, reproduced the anomaly in three dimensions using the conformal completion technique of Ref. ~\refcite{Ashtekar:1999jx}. It will be interesting to see why in this approach the anomaly doesn't show up in higher dimensions and also how the anomaly is reflected in the covariant phase space Hamiltonian formalism.

\section*{Acknowledgments} 
AM would like to express his deepest gratitude to Prof. Enric Verdaguer for his kind hospitality at the Department de F\'{i}sica Fonamental de la Universitat de Barcelona where this work was completed.

\end{document}